\documentclass[aps,prb,twocolumn,superscriptaddress,amsmath]{revtex4-1}

\usepackage{graphicx}
\usepackage{bm}

\begin{document}

\title{Nonlocal transport and heating in superconductors under dual-bias conditions}

\author{S. Kolenda}
\author{M. J. Wolf}
\affiliation{Institut f\"ur Nanotechnologie, Karlsruher Institut f\"ur Technologie, Karlsruhe, Germany}
\author{D. S. Golubev}
\affiliation{Institut f\"ur Nanotechnologie, Karlsruher Institut f\"ur Technologie, Karlsruhe, Germany}
\affiliation{Low Temperature Laboratory (OVLL), Aalto University School of Science, P.O. Box 13500, 00076 Aalto, Finland}
\author{A. D. Zaikin}
\affiliation{Institut f\"ur Nanotechnologie, Karlsruher Institut f\"ur Technologie, Karlsruhe, Germany}
\affiliation{I.E.Tamm Department of Theoretical Physics, P.N.Lebedev Physics Institute, 119991 Moscow, Russia}
\affiliation{Laboratory of Cryogenic Nanoelectronics, Nizhny Novgorod State Technical University, 603950 Nizhny Novgorod, Russia}
\author{D. Beckmann}
\affiliation{Institut f\"ur Nanotechnologie, Karlsruher Institut f\"ur Technologie, Karlsruhe, Germany}
\email[e-mail address: ]{detlef.beckmann@kit.edu}

\date{\today}

\begin{abstract}
We report on an experimental and theoretical study of nonlocal transport in superconductor hybrid structures, where two normal-metal leads are attached to a central superconducting wire. As a function of voltage bias applied to both normal-metal electrodes, we find surprisingly large nonlocal conductance signals, almost of the same magnitude as the local conductance. We demonstrate that these signals are the result of strong heating of the superconducting wire, and that under symmetric bias conditions, heating mimics the effect of Cooper pair splitting.
\end{abstract}


\maketitle

\section{Introduction}

In hybrid proximity structures consisting of a normal metal and a superconductor (NS) electrons can be converted into Cooper pairs.
Depending on the electron energy this conversion is provided by different physical mechanisms. Electrons with overgap energies may easily penetrate from a normal metal deep into a superconductor causing electron-hole branch imbalance \cite{tinkham1972,*tinkham1972b} which relaxes at macroscopic distances from the NS interface. In contrast, electrons with subgap energies penetrate into a superconductor by the mechanism of Andreev reflection (AR).\cite{andreev1964}
In this case an electron propagating in a normal metal may enter a superconductor only at a rather short distance (of order of
the superconducting coherence length $\xi$) forming a Cooper pair
together with another electron taken from the same normal metal. At sufficiently low energies this Andreev reflection mechanism is
responsible for dissipative charge transfer across NS interfaces.\cite{blonder1982}

In multi-terminal hybrid proximity structures, such as NSN systems,
the physics of low energy electron transport becomes much richer as it also includes coherent {\it non-local} effects. Provided the
superconductor size (i.e. the distance between two NS interfaces) is comparable with (or smaller than) $\xi$, two extra charge transfer mechanisms gain importance. One of them is the so-called elastic cotunneling (EC), i.e. direct transfer of subgap electrons
between two N-metals through a superconductor. Another
mechanism is crossed (or non-local) Andreev reflection \cite{byers1995,deutscher2000} (CAR).  In contrast to local AR, here a Cooper pair
is formed by two electrons penetrating into a superconductor from two
different N-terminals. This mechanism essentially
influences non-local charge transport in hybrid NSN systems. Furthermore, employing the phenomenon of CAR one
can provide a direct experimental realization of entanglement between
electrons in different normal terminals. In other words, three-terminal NSN devices can effectively act as Cooper pair splitters.\cite{lesovik2001,recher2001,hofstetter2009,herrmann2010}

Both experimental\cite{beckmann2004,beckmann2007,russo2005,cadden2006,kleine2009,almog2009,kleine2010,brauer2010,wei2010,kaviraj2011} and theoretical\cite{falci2001,bignon2004,brinkman2006,morten2006,kalenkov2007c,duhot2007,golubev2007,yeyati2007,kalenkov2007,kalenkov2008,golubev2009b,golubev2009,bergeret2009,golubev2010,freyn2010}
(see also further references therein) investigations of dissipative electron transport and non-local shot noise in three-terminal NSN structures revealed a rich variety of non-trivial
features. For instance, in the tunneling limit and at $T\to 0$ EC and CAR contributions to non-local
conductance exactly cancel each other,\cite{falci2001} thus
leaving no possibility to experimentally test the effect of CAR
in transport experiments in this limit. Splitting the contributions
of EC and CAR becomes possible either at higher interface transmissions
\cite{kalenkov2007c,kalenkov2007} or by applying an external ac field \cite{golubev2009b} or, else, by studying non-local shot noise.\cite{bignon2004,wei2010,kaviraj2011,golubev2010}

Further interesting features emerge in the presence of disorder. In this case an interplay
between CAR, quantum interference of electrons and non-local charge imbalance dominates the behavior of diffusive NSN systems \cite{golubev2009} and, for instance, may yield strong enhancement of non-local conductance in the low energy limit. The effect of disorder needs to be taken into account for a quantitative interpretation of the experiments.\cite{cadden2006,kleine2009,almog2009}

Non-trivial physics also emerges from an interplay between CAR and
Coulomb interaction. E.g., interactions lift the exact
cancellation of EC and CAR contributions to the non-local differential conductance already in the lowest order in tunneling.\cite{yeyati2007} This conductance is predicted to have an S-like shape and can turn negative at non-zero bias.\cite{golubev2010} Furthermore, one can prove \cite{golubev2010} that there exists a fundamental relation between Coulomb effects and non-local shot noise in NSN structures which can be directly tested in future experiments.

In this work we will explore yet another physical effect which -- along
with the above mentioned ones -- can essentially influence the behavior of three-terminal NSN proximity structures. Namely, we will demonstrate -- both experimentally and theoretically -- that non-local transport properties of such structures can be strongly affected (or even dominated) by heating.
It is important to emphasize that under certain conditions heating can mimic both the effect of CAR and Cooper pair splitting.
Thus, it is in general mandatory to account for heating effects while analyzing non-local phenomena in multi-terminal proximity structures.

The structure of our paper is as follows. In Sec. II we describe our NSN samples as well as the key aspects of our experiments. In Sec. III we
present our experimental results demonstrating an importance of heating
effects for non-local electron transport in the structures under consideration. Theoretical model aimed to quantitatively explain our
experimental observations is worked out in Sec. IV. In Sec. V we
make use of this model demonstrating a good agreement between theory
and experiment without involving any fit parameters. Technical details of our derivation of the Coulomb correction to the non-local conductance
are displayed in Appendix.

\section{Samples and Experiment}

\begin{figure}[bt]
\includegraphics[width=\columnwidth]{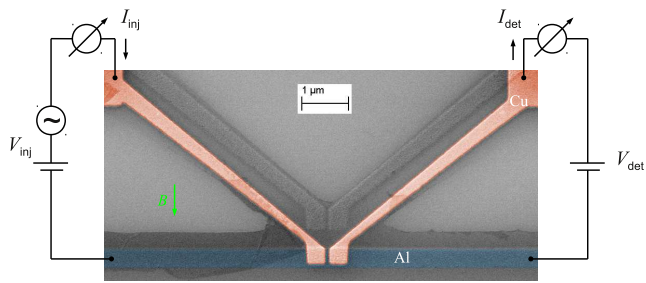}
\caption{\label{fig_sem}(color online) False color scanning electron microscopy image of a section of sample B, together with the measurement scheme.}
\end{figure}

Our samples consist of a central superconducting wire, with several copper wires attached by tunnel junctions of normal-state conductance $G_\mathrm{T}$. The junctions are connected to the reservoirs by long normal-metal wires, which introduce a series resistance $r^N$ to create an Ohmic enviroment for the junctions. Figure~\ref{fig_sem} shows a false-color scanning electron microscopy image of one of the samples as well as a scheme of the measurement setup.  The samples were fabricated by standard e-beam lithography and shadow evaporation techniques. In a first step, the superconducting aluminum wire of thickness $t_\mathrm{Al} = 20$~nm  was created. The  aluminum wire was oxydized \textit{in situ} to form a thin but pinhole-free tunnel barrier by exposing it to about $0.5$~Pa of pure oxygen for a few minutes. After the oxidation, copper of  thickness $t_\mathrm{Cu}=15-30$~nm was evaporated under a different angle to form the tunnel junctions.  We investigated samples with two closely-spaced tunnel junctions as shown in Fig.~\ref{fig_sem}, as well as one sample with six junctions to investigate the dependence of non-local transport on the contact distance $d$ (not shown).  An overview of the sample parameters is given in Table~\ref{tab_sample_properties}.

All measurements were performed in a dilution refrigerator at temperatures down to $T = 50$~mK. A magnetic field $B$ could be applied in the substrate plane perpendicular to the aluminum wire, as indicated in Fig.~\ref{fig_sem}. A voltage $V_\mathrm{inj}$ consisting of a dc bias and a low-frequency ac excitation was applied to one tunnel contact, called injector, and the ac part of the resulting current $I_\mathrm{inj}$ was measured by standard lock-in techniques to obtain the local conductance $G_\mathrm{loc} = dI_\mathrm{inj} / dV_\mathrm{inj}$. Simultaneously, the ac current $I_\mathrm{det}$ through the second contact, held at fixed bias voltage $V_\mathrm{det}$, was measured to determine the nonlocal conductance $G_{\textrm{nl}} = dI_\mathrm{det} / dV_\mathrm{inj}$.

Our key experimental results obtained in this way are outlined in the next section.

\begin{table}[bt] 
\caption{\label{tab_sample_properties}Overview of sample properties. Number of  tunnel junctions, normal-state tunnel conductance $G_\textrm{T}$, series resistance $r^N$, thickness $t_\textrm{Cu}$ and length $l_\textrm{Cu}$ of the copper wire.}
\begin{ruledtabular}
\begin{tabular}{lccccc}
        &          & $G_\textrm{T}$ & $r^N$ & $t_\textrm{Cu}$ & $l_\textrm{Cu}$ \\
Sample  & junctions & (mS)          & ($\Omega$)     & (nm)            & ($\mu$m)        \\ \hline
A       & 2        &  $3.2$         & $240$          & 15              & 15              \\
B       & 2        &  $4$           & $15$           & 30              & 4.8             \\
C       & 6        &  $1$           & $13$           & 30              & 3               \\
\end{tabular}
\end{ruledtabular}
\end{table}

\section{Experimental results}

\begin{figure}[bt]
\includegraphics[width=\columnwidth]{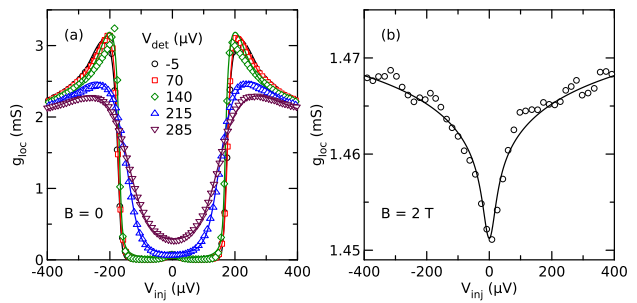}
\caption{\label{fig_loc}
(color online)
Local differential conductance $g_\mathrm{loc}$ of a junction of sample A as a function of injector bias $V_\mathrm{inj}$ at $T=20~\mathrm{mK}$. (a) data for different detector bias voltages $V_\mathrm{det}$ and at zero field in the superconducting state. (b) data at $B=2~\mathrm{T}$ in the normal state.}
\end{figure}

Figure~\ref{fig_loc}(a) shows the local conductance of a junction of sample A as a function of injector bias $V_\mathrm{inj}$ for different detector bias $V_\mathrm{det}$ at low temperature in the superconducting state. The data exhibit a well-defined energy gap $\Delta\approx 180~\mathrm{\mu eV}$ and coherence peaks for $|eV_\mathrm{det}|<\Delta$, and an increased broadening for $|eV_\mathrm{det}|>\Delta$. In Figure~\ref{fig_loc}(b), we also show the conductance at high magnetic field in the normal state. Here, a dip due dynamical Coulomb blockade is observed.

In order to fit the local conductance, we model the density of states $\nu(E)$ in the superconductor including a phenomenological life-time broadening parameter $\Gamma$ (the so-called Dynes parameter \cite{dynes1978}),
\begin{equation}
 \nu(E)=\mathrm{Re}\left(\frac{E+i \Gamma}{\sqrt{(E+i \Gamma)^2-\Delta^2}}\right)
\end{equation}
where $E$ is the quasiparticle energy and $\Delta$ is the gap. The current through the tunnel junction is then given by
\begin{equation}
  I_\mathrm{T}(V_\mathrm{T}) = \frac{G_\mathrm{T}}{e} \int \nu(E) \left(f_0(E)-f_0(E+eV_\mathrm{T})\right) dE,
  \label{eqn_gloc}
\end{equation}
where $V_\mathrm{T}$ is the voltage across the junction and $f_0$ is the Fermi function. For the samples with long copper wire, the series resistance $r^N_\mathrm{inj}$ is of similar magnitude as the junction resistance $1/G_\mathrm{T}$, and we cannot neglect the voltage drop across $r^N_\mathrm{inj}$. The actual voltage across the junction is therefore $V_\mathrm{T}=V_\mathrm{inj}-r^N_\mathrm{inj} I_\mathrm{inj}$, and we solve the implicit equation
\begin{equation}
  I_\mathrm{inj}=I_\mathrm{T}(V_\mathrm{inj}-r^N_\mathrm{inj} I_\mathrm{inj})
\end{equation}
for $I_\mathrm{inj}$ to fit the data. We thus have $G_\mathrm{T}$, $r^N_\mathrm{inj}$, $\Delta$, $\Gamma$ and the temperature as fit parameters. We denote the temperature from these fits by $T_\mathrm{N}$, since it actually describes the smearing of the Fermi distribution in the normal metal. While $\Gamma$ and $T_\mathrm{N}$ describe a similar broadening of the conductance features, we found that both had to be adjusted to give a good fit of the data. Fits to this model are shown as lines in Fig.~\ref{fig_loc}(a). We proceeded by first fitting a trace at large detector bias, and then kept $G_\mathrm{T}=3.2~\mathrm{mS}$ and $r^N_\mathrm{inj}=240~\Omega$ fixed for all other fits. The high field data are fit with the standard model of dynamical Coulomb blockade\cite{devoret1990,schoen1990,ingold1992}, shown as a line in Fig.~\ref{fig_loc}(b). For the latter fit, we kept the junction conductance fixed to its value in the superconducting state, and fit the series resistance $r^N_\mathrm{inj}=370~\Omega$ as well as the effective impedance $R^E=80~\Omega$ of the electromagnetic enviroment. The fact that $r^N_\mathrm{inj}$ in the normal state is larger than in the superconducting state reflects that the resistance of the aluminum wire now also appears in series with the junction. On the other hand, $R^E<r^N_\mathrm{inj}$ indicates that only a fraction of the series resistance actually affects dynamical Coulomb blockade, which probes the environment at high frequencies $h\nu\approx eV$. A similar fit for sample B with the shorter Cu wire yields $R^E=40~\Omega$.

\begin{figure}[bt]
\includegraphics[width=\columnwidth]{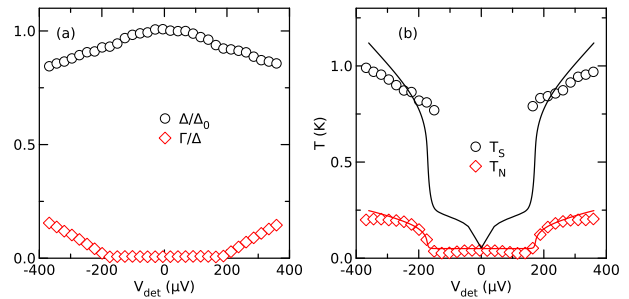}
\caption{\label{fig_parameters}
(color online) Fit parameters of the local conductance of the sample A.
(a) normalized energy gap $\Delta$ and life-time broading parameter $\Gamma$ as a function of detector bias $V_\mathrm{det}$.
(b) effective temperatures $T_\mathrm{S}$ and $T_\mathrm{N}$ as a function of detector bias $V_\mathrm{det}$ derived from the fits (symbols), and model predictions (lines).}
\end{figure}

The parameters extracted from the fits in the superconducting state are shown in Fig.~\ref{fig_parameters} as a function of detector bias $V_\mathrm{det}$. Figure~\ref{fig_parameters}(a) shows the gap $\Delta$ normalized to its value $\Delta_0=180~\mathrm{\mu eV}$ at zero bias, as well as the normalized life-time broadening parameter $\Gamma$. $\Delta$ decreases by about 15~\% with increasing bias, whereas $\Gamma$ remains about zero for $V_\mathrm{det}<\Delta_0/e$, and sharply increases as soon as the bias exceeds the gap. The effective temperature $T_\mathrm{N}$ behaves in a similar way as $\Gamma$, as seen in Fig.~\ref{fig_parameters}(b). It remains close to the bath temperature $T=50~\mathrm{mK}$ below the gap, and then quickly increases to $T_\mathrm{N}\approx 200~\mathrm{mK}$. To estimate the effective temperature $T_\mathrm{S}$ of the quasiparticles in the superconductor, we have inverted the BCS temperature dependence of the gap to relate the decrease of $\Delta$ as a function of bias to an increase of $T_\mathrm{S}$. Since $\Delta(T)$ is almost flat at low temperatures, the result of this inversion is not very reliable for small deviations of $\Delta$ from $\Delta_0$. We therefore only plot the resulting $T_\mathrm{S}$ for large bias in Fig.~\ref{fig_parameters}(b). As can be seen,  $T_\mathrm{S}\approx  1~\mathrm{K}\gg T_\mathrm{N}$. This results is reasonable since the electrons in the normal metal are heated indirectly by the quasiparticles in the superconductor.

\begin{figure}[bt]
\includegraphics[width=\columnwidth]{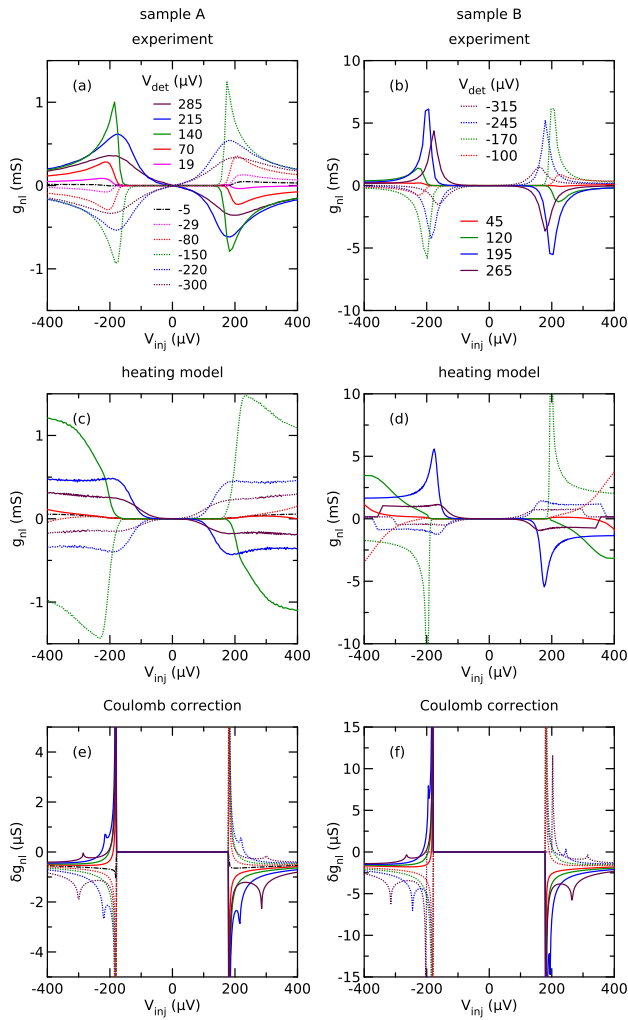}
\caption{\label{fig_nonlocal}
(color online)
Nonlocal differential conductance $g_\mathrm{nl}$ of a pair of junctions of samples A (a) and B (b) as a function of injector bias $V_\mathrm{inj}$ for different detector bias voltages $V_\mathrm{det}$. (c) and (d): corresponding predictions of the heating model discribed in Sec. \ref{sec:model}. (e) and (f): Coulomb correction predicted by eq. (\ref{dIdVCol}). Note the different scale.}
\end{figure}

Figure~\ref{fig_nonlocal} shows the nonlocal conductance for samples A and B. For (nearly) zero detector bias, we observe a signal of a few ten $\mathrm{\mu S}$. A signal of this magnitude is expected due to charge imbalance, as described in detail in a previous publication.\cite{huebler2010} The signal due to charge imbalance is an even function of injector bias, see  Ref.~\onlinecite{huebler2012b}. For finite detector bias, we observe two peaks in the nonlocal conductance. These exceed the charge imbalance signal by orders of magnitude, and are odd functions of both injector and detector bias. The peaks are relatively sharp and initially increase for $|eV_\mathrm{det}|<\Delta$. For $|eV_\mathrm{det}|>\Delta$, the peaks broaden and decrease considerably, much like the coherence peaks in the local conductance shown in Figure~\ref{fig_loc}(a). The nonlocal conductance is negative if the sign of the injector and detector bias is the same, and positive otherwise. Symmetric bias conditions $V_\mathrm{inj}=V_\mathrm{det}$ are the typical operating point for Cooper-pair splitter devices. Under these conditions, we observe negative nonlocal conductance, the same as one would find for crossed Andreev reflection.

\begin{figure}[bt]
\includegraphics[width=\columnwidth]{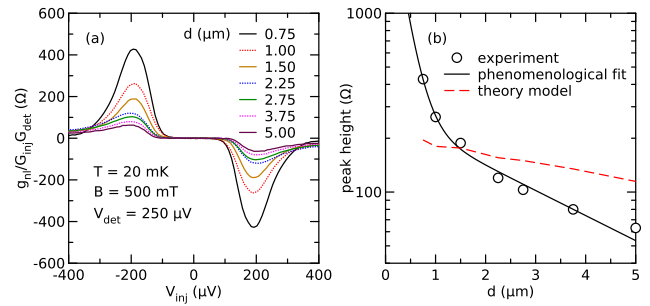}
\caption{\label{fig_distance}
(color online)
(a)  Normalized nonlocal conductance $g_\mathrm{nl}/G_\mathrm{inj}G_\mathrm{det}$ for fixed detector bias $V_\mathrm{det}=250$ $\mu$V 
for different contact distances $d$. (b) peak height as a function of contact distance $d$. The solid line is a phenomenological fit to a 
two-scale exponential decay, with relaxation lengths $\lambda_1=0.23~\mathrm{\mu m}$ and $\lambda_2=3.1~\mathrm{\mu m}$. The dashed line 
indicates the model prediction (see text).}
\end{figure}

Finally, in Fig.~\ref{fig_distance}(a) we show the dependence of the nonlocal conductance on contact distance $d$ for a fixed detector bias measured 
on sample C. The peaks described previously decrease slowly upon increasing $d$, with little change in the overall shape. 
The maximum conductance is plotted on a semi-logarithmic scale as a function of  $d$ in Fig.~\ref{fig_distance}(b). 
The decay can be fit by the sum of two exponentials, with relaxation lengths $\lambda_1=0.23~\mathrm{\mu m}$ and  $\lambda_2=3.1~\mathrm{\mu m}$. 
The slow decay over a scale of several microns is compatible with nonequilibrium quasiparticle transport.

\section{Non-local conductance and heating: Theoretical model}\label{sec:model}

The behavior of the non-local conductance presented in Figs. 4 and 5 qualitatively resembles that in the presence of Coulomb effects \cite{golubev2010}. On the other hand, our data demonstrate that - at least at sufficiently high voltages -- heating effects are essential and, hence, should also be taken into account. The task at hand is to formulate a theoretical model which would adequately describe our system in the presence of both Coulomb interactions and heating effects.

In order to construct a complete theory of non-equilibrium heat transport in $SN$ proximity structures it is in general necessary to employ the Keldysh technique and to work out a solution of inhomogeneous Usadel equations \cite{belzig1999}, see, e.g., Ref. \onlinecite{virtanen2007} for a review. The corresponding analysis turns out to be rather involved. For instance, substantial technical complications are due to the fact that in our problem the superconducting gap  may acquire a significant coordinate dependence $\Delta(\bm{r})$ induced both by the proximity effect and by an inhomogeneous temperature profile inside the wire.
Yet another complication has to do with the presence of electron-electron scattering, which leads to an effective
equilibration of the quasiparticle distribution function at a certain length scale. 

In order to proceed, below we will employ a simple  
model of non-local charge and heat transport through the structure depicted in Fig. 1.
Within this model we will ignore the proximity effect assuming the resistances of the tunnel junctions to be sufficiently high. 
Then the coordinate dependence of $\Delta$ may become important only at sufficiently
strong overheating, i.e. provided the heat transport properties of a  superconductor already resemble those of a normal metal. Our model does take into account the coordinate
dependence of $\Delta$, but disregards Andreev reflection of quasiparticles associated with it. Therefore we expect it to be more accurate at relatively small overheating (low bias)
and less accurate, but still qualitatively correct,
at strong overheating (high bias). Within our model we will also
assume that the electron-electron energy relaxation length is the shortest length scale in our problem 
and that the electron distribution function coincides with the Fermi function with the local electron temperature. 
The latter may deviate from the temperature of the phonon subsystem, which we assume to be the same as the base temperature of the cryostat.
In order to justify the above assumption one can make use simple theoretical estimates of the electron-electron relaxation length at high energies 
and/or just quote earlier experiments with normal wires \cite{pothier1997}, in which this length -- under the conditions
similar to ours -- was found to be of the order of or shorter than 1 $\mu$m. This length scale is significantly shorter than the length of normal wires
in our setup. We expect similar values of the relaxation length for
strongly excited quasiparticles in aluminum, which give the main contribution to the signal at high bias. Thus
the same arguments may be applied to the superconducting wire as well.   
As we will demonstrate, our simple model rather accurately describes the properties of the system under consideration.

Under the conditions outlined above the current through the detector junction can be expressed in the form
\begin{eqnarray}
I_{\rm det} &=& -I^{\rm det}_{\rm T}(V_1,T_{\rm det},T_{\rm det}^S)
+ I^{\rm det}_{\rm CI}(V_1,T_{\rm det},T_{\rm det}^S)
\nonumber\\ &&
+\, I^{\rm nl}_{\rm CI}(V_2,T_{\rm inj},T_{\rm inj}^S)
+\delta I_{\rm Col}^{\rm det}(V_1,V_2).
\label{current_1}
\end{eqnarray}
Here $V_1,V_2$ are the voltage drops across the detector and injector junctions respectively.
They are related to the potentials $V_{\rm det}$ and $V_{\rm inj}$ (see Fig. 1) as follows
\begin{eqnarray}
V_{\rm det} &=& V_1 + I_{\rm det}r^N_{\rm det},
\label{Vdet}\\
V_{\rm inj} &=& V_2 + I_{\rm inj}r^N_{\rm inj},
\label{Vinj}
\end{eqnarray}
where $r^N_{\rm det}$ and $r^N_{\rm inj}$ are the resistances of the normal wires
attached, respectively, to the detector and injector junctions.
The voltage $V_2$ actually
coincides with the voltage $V_{\rm T}$ already introduced in Eq. (\ref{eqn_gloc}).

The current of the detector  (\ref{current_1}) is the sum of four contributions. The first one, $-I^{\rm det}_{\rm T}(V_1,T_{\rm det},T_{\rm det}^S)$,  is the standard tunneling current between normal and superconducting wires
defined in Eq. (\ref{eqn_gloc}). The minus sign in front of this contribution is due to the adopted sign convention, see Fig. 1.
The second and the third contributions arise from the charge imbalance (CI) induced by non-equilibrium quasiparticles injected into the superconducting wire respectively through the detector junction and the injector,
\begin{eqnarray}
&& I^{\rm det}_{\rm CI}(V_1,T_{\rm det},T_{\rm det}^S) =
\frac{\left(G_{\rm T}^{\rm det}\right)^2r_Lr_R}{e(r_L+r_R)}\int dE\,
\nonumber\\ && \times\,
\theta(|E|-\Delta(T^S_{\rm det}))\big[ f_{\rm det}(E-eV_1,T_{\rm det})
- f_S(E,T_{\rm det}^S) \big],
\nonumber\\
\end{eqnarray}
\begin{eqnarray}
&& I^{\rm nl}_{\rm CI}(V_2,T_{\rm inj},T_{\rm inj}^S) =
\frac{G_{\rm nl}^{(0)}}{e}\int dE\, \theta(|E|-\Delta(T^S_{\rm inj}))
\nonumber\\ && \times\,
\big[ f_{\rm inj}(E-eV_2,T_{\rm inj})
- f_S(E,T_{\rm inj}^S) \big],
\label{Inl}
\end{eqnarray}
and, finally, the fourth term, $\delta I_{\rm Col}^{\rm det}(V_1,V_2)$, is the Coulomb interaction correction \cite{golubev2010} derived in Appendix.

In the above expressions we introduced the following parameters:
$r_L$ and $r_R$ are the left and right normal state resistances of the segments of the superconducting wire between the corresponding junctions
and the bulk leads,
$f_{\rm det}(E,T_{\rm det})=1/[1+e^{E/T_{\rm det}}]$
and
$f_S(E,T_S)=1/[1+e^{E/T_S}]$
are the quasiparticle distribution functions respectively in the normal lead attached to the detector junction with temperature
$T_{\rm det}$ and  in the superconductor with the local temperature $T_S$, $G_{\rm nl}^{(0)}\sim G_{\rm T}^{\rm det}G_{\rm T}^{\rm inj} r_Lr_R/(r_L+r_R)$
is the normal state value of the non-local conductance unaffected by Coulomb interaction.

According to Eqs. (\ref{current_1}) and (\ref{Vinj}) the non-local conductance of the system reads
\begin{eqnarray}
G_{\rm nl} &=& \frac{d I_{\rm det}}{d V_{\rm inj}}
\nonumber\\
&=& \frac{\frac{\partial I_{\rm det}}{\partial V_2}
+\frac{\partial I_{\rm det}}{\partial T_{\rm det}}\frac{\partial T_{\rm det}}{\partial V_2}
+\frac{\partial I_{\rm det}}{\partial T_{\rm det}^S}\frac{\partial T_{\rm det}^S}{\partial V_2}}
{1+\left(\frac{\partial I_{\rm inj}}{\partial V_2}
+\frac{\partial I_{\rm inj}}{\partial T_{\rm inj}}\frac{\partial T_{\rm inj}}{\partial V_2}
+\frac{\partial I_{\rm inj}}{\partial T_{\rm inj}^S}\frac{\partial T_{\rm inj}^S}{\partial V_2}\right) r^N_{\rm inj}}.
\label{Gnl}
\end{eqnarray}
This formula expresses $G_{\rm nl}$ as a function of the voltages $V_1$ and $V_2$. In order to re-formulate it in terms of experimentally accessible voltages $V_{\rm det}$ and $V_{\rm inj}$, Eq. (\ref{Gnl})
should be employed in combination with Eqs. (\ref{Vdet}), (\ref{Vinj}).

The non-local conductance (\ref{Gnl}) contains two types of derivatives. The derivatives
${\partial I_{\rm det}}/{\partial V_2}$ and ${\partial I_{\rm inj}}/{\partial V_2}$ remain
finite even in equilibrium when the sample is well cooled and the temperature values in all electrodes
do not depend on the bias voltages. The terms containing the derivatives over temperature account for the heating effect. It turns out that 
these terms give the dominant contribution to the non-local conductance in our samples.

As compared to the above terms, the Coulomb correction to the non-local conductance $\partial\,\delta I_{\rm Col}^{\rm det}(V_1,V_2)/\partial V_2$ remains small
and can be disregarded for the structures under consideration.
In order to see that, let us set $T \to 0$ and choose $e|V_1|,e|V_2|>\Delta $. In this case the derivative
${\partial\,\delta I_{\rm Col}^{\rm det}(V_1,V_2)}/{\partial V_2}$  may be expressed in a relatively simple form (see Appendix for details)
\begin{eqnarray}
&& \frac{\partial\,\delta I_{\rm Col}^{\rm det}(V_1,V_2)}{\partial V_2} \approx
-\frac{G_{\rm nl}^{(0)}}{g_{\rm det}^E}\ln\bigg(1+\frac{V_0^2}{( V_1 - V_2 )^2}\bigg)
\nonumber\\ &&
-\, \frac{G_{\rm nl}^{(0)}}{2g_{\rm det}^E}\frac{\Delta^2}{e^2V^2_2-\Delta^2}
\bigg[ \ln\bigg(1+\frac{V_0^2}{( V_1 - V_2 )^2}\bigg)
\nonumber\\ &&
-\, \ln\bigg(1+\frac{V_0^2}{( V_1 + V_2 )^2}\bigg) \bigg],
\label{dIdVCol}
\end{eqnarray}
where we defined the high voltage cutoff $eV_0\sim 1/\tau_{RC}$ determined by the inversed effective $RC-$time of our system
and introduced the dimensionless conductance $g_{\rm det}^E= 2\pi /e^2 R^E_{\rm det}$ of the electromagnetic environment ``seen'' 
by the detector. For our samples, we typically have $g_{\rm det}^E\gg 1$. Thus, the Coulomb correction $\propto 1/g_{\rm det}^E$ remains small 
except, perhaps, an immediate vicinity of the gap voltage $eV_1=\Delta$, see Fig. 4e-f. This observation allows us to ignore the Coulomb interaction 
correction in our further consideration.

Our next step is to find the dependence of the temperatures $T_{\rm inj},T_{\rm det}$, $T^S_{\rm inj},T^S_{\rm det}$ on the
bias voltages $V_{\rm inj}$ and $V_{\rm det}$. For this purpose
it will be necessary to solve the corresponding heat transport equations.

Let us first consider the normal lead attached to the detector.
We will approximately treat it as a thin quasi-one-dimensional wire.
The equation describing the heat transport along the wire reads
\begin{eqnarray}
P_{\rm det} = -\Sigma S_{\rm det} \int_0^x dx'\, \left( T_{\rm det}^5(x') - T_0^5 \right)
\nonumber\\
+ \, \frac{I_{\rm det}^2 x}{\sigma S_{\rm det}} + \frac{\pi^2 \sigma S_{\rm det}}{6e^2} \frac{d}{dx} T_{\rm det}^2(x),
\label{heat_balance_1}
\end{eqnarray}
where we defined the coordinate $x$ along the wire
and assumed that the detector junction is located at $x=0$.
The quantity $P_{\rm det}$ denotes the heat power extracted from the normal wire or, equivalently, the cooling power of a detector wire. 
This quantity is given by the integral \cite{giazotto2006}
\begin{eqnarray}
P_{\rm det}(V_1,T_{\rm det},T_{\rm det}^S) = \frac{G_{\rm T}^{\rm det}}{e}\int dE\, \nu(E)(E-eV_1)
\nonumber\\ \times\,
[ f_{\rm det}(E-eV_1,T_{\rm det})- f_S(E,T_{\rm det}^S) ],
\label{P1}
\end{eqnarray}
which remains positive at $eV_1\lesssim \Delta$ and $T_{\rm det}=T_{\rm det}^S$ and turns negative in the
high bias regime $eV_1\gtrsim\Delta$, $T_{\rm det}<T_{\rm det}^S$.
The first term in the right hand side of Eq. (\ref{heat_balance_1}) describes
the heat current from the electron subsystem into the phonon one,
the material parameter $\Sigma$ characterizes the electron-phonon coupling strength,
and $S_{\rm det}$ stands for the cross sectional area of the detector normal wire.
The second term in Eq. (\ref{heat_balance_1}) describes the Joule heating of the wire by the current,
and the last term is the heat current flowing along the wire and leaking into the outer bulk electrode.
According to the Wiedemann - Franz law this heat current
is proportional to the conductivity of the normal wire $\sigma$.
Thus, Eq. (\ref{heat_balance_1}) implies that the power generated in
the biased detector junction is partially dissipated in the phonon subsystem and partially carried away along the wire.

Finally, Eq. (\ref{heat_balance_1}) should be supplemented by the boundary
conditions
\begin{eqnarray}
T_{\rm det}(0)=T_{\rm det},\;\; T_{\rm det}(L_{\rm det})=T_0,
\end{eqnarray}
where $L_{\rm det}$ is the length of the detector normal wire.
Here we assumed that at $x=L_{\rm det}$ the wire is coupled
to a bulk metallic lead kept at the base temperature $T_0$.
The heat transport in the normal wire attached to the injector is described by Eq. (\ref{heat_balance_1}) with interchanged indices.

In  order to fit our data, we numerically solved the heat balance equation (\ref{heat_balance_1}).
For the parameters of our samples we verified that
the term responsible for electron-phonon interactions
may be omitted provided the wire is short enough, i.e. $L_{\rm det}\ll L_{\rm e-ph}$, where
\begin{eqnarray}
L_{\rm e-ph}=\sqrt{\frac{\pi^2\sigma}{6e^2\Sigma T_{\rm det}^3}}
\label{lambda}
\end{eqnarray}
is the electron-phonon relaxation length.
For the copper wire one has \cite{giazotto2006} $\Sigma\approx 2$ nW/$\mu$m$^3$K$^5$.
Combining this value with
with the conductivity of our copper leads, $\sigma\approx 45$ $(\mu\Omega\, {\rm m})^{-1}$, we find
$L_{\rm e-ph}= 1500$ $\mu$m at $T=50$ mK and
$L_{\rm e-ph}= 17$ $\mu$m  at $T=1$ K.
Thus, the electron-phonon relaxation length indeed exceeds the
length of the normal wire in both our samples in the whole range of temperatures relevant for our experiment.

Hence, we can safely omit the electron-phonon term from the differential equation (\ref{heat_balance_1}).
With this in mind one can easily integrate this equation reducing it to the algebraic one
\begin{eqnarray}
\frac{\pi^2\left(T_{\rm det}^2 - T_0^2\right)}{6e^2r_{\rm det}^N}+P_{\rm det}-\frac{I^2_{\rm det}r_{\rm det}^N}{2}=0,
\label{T1}
\end{eqnarray}
where $r_{\rm det}^N=L_{\rm det}/\sigma S_{\rm det}$ is the resistance of the detector normal wire.
Similarly, for the injector normal wire one finds
\begin{eqnarray}
\frac{\pi^2\left(T_{\rm inj}^2 -T_0^2\right)}{6e^2r_{\rm inj}^N}+P_{\rm inj}-\frac{I^2_{\rm inj}r_{\rm inj}^N}{2}=0.
\label{T2}
\end{eqnarray}

We now turn to the heat transport equation in the superconducting wire.
In this case the heat current from quasiparticles to phonons is
in general defined by a rather complicated double integral.
Here we will disregard the corresponding term in our heat transport equation from the very beginning
assuming that overheating of our superconducting wire remains
not too strong.
This approximation requires that the superconducting wire length $L_S$ is smaller than the electron-phonon relaxation length in the superconductor $L_{\rm e-ph}^S$, i.e.
\begin{eqnarray}
L_S \ll \sqrt{\frac{\pi^2\sigma_S}{6e^2\Sigma_S T_S^3}},
\label{condition_S}
\end{eqnarray}
where $\sigma_S$ and $\Sigma_S$ are the conductivity and the electron-phonon coupling parameter
in the normal state of the superconductor.
We find that for our samples the condition (\ref{condition_S}) is satisfied in the range
of voltages $e|V_{\rm inj}|,e|V_{\rm det}|\lesssim 1.5\Delta$, but may be not fulfilled at higher voltages.

In order to further simplify our model we also assume that
the detector and injector junctions are located close enough to each other, meaning that the temperature values on their superconducting
sides are the same, i.e. $T_{\rm det}^S=T_{\rm inj}^S=T_S$.
Under these conditions we can write the heat transport equation in the form
\begin{eqnarray}
P^+ &=& P_{\rm qp}(T(x)),\;\; x>0,
\nonumber\\
-P^- &=& P_{\rm qp}(T(x)),\;\; x<0
\nonumber\\
P^+ + P^- &=& I_{\rm inj}V_{2} + P_{\rm inj} + I_{\rm det}V_{1} + P_{\rm det}.
\label{heat_balance_S}
\end{eqnarray}
Here we assumed that both junctions are in the vicinity of the point $x=0$,
introduced the temperature of the superconductors at the point $x$, $T(x)$,
the heat current $P^+$ flowing to the direction $x>0$ and the heat
current $P^-$ flowing in the opposite direction $x<0$.
The combination $I_{\rm inj}V_{2} + P_{\rm inj} + I_{\rm det}V_{1} + P_{\rm det}$
is the total heat power injected into the superconductor by both tunnel junctions.
It is given by the sum of the Joule heating by both junctions, $I_{\rm inj}V_{2} + I_{\rm det}V_{1}$, and the total
cooling power of both normal wires, $P_{\rm inj}+P_{\rm det}$.
The boundary conditions for Eqs. (\ref{heat_balance_S}) read
\begin{eqnarray}
T(0)=T_S,\;\; T(-L_{S1})=T_0,\;\; T_S(L_{S2})=T_0,
\end{eqnarray}
where $L_{S1}$ and $L_{S2}$ are the lengths of the wire segments on both sides of the junctions.

Next, $P_{\rm qp}(T(x))$ is the quasiparticle heat current in the superconductor in presence of the
temperature gradient. It reads
\begin{eqnarray}
P_{\rm qp}(T(x)) = -\kappa(T(x))S_S\frac{dT(x)}{dx},
\end{eqnarray}
where $S_S$ is the cross sectional area of the superconducting wire and the heat conductivity of the superconductor is defined as\cite{bardeen1959}
\begin{eqnarray}
&& \kappa(T) = \frac{2\sigma_S}{e^2} T\bigg[ 2F\left(\frac{\Delta}{T}\right)
+\frac{2\Delta}{T}\ln\left(1+e^{-\Delta/T}\right)
\nonumber\\ &&
+\, \frac{\Delta^2}{T^2\left(1+e^{\Delta/T}\right)}\bigg],\;\;
F(x) = \int_0^\infty dz\frac{z}{1+e^{z+x}}.
\label{kappa}
\end{eqnarray}

Eq. (\ref{heat_balance_S}) can be integrated in exactly the same way as Eq. (\ref{heat_balance_1}). As a result, we arrive at the following algebraic equation
\begin{eqnarray}
{\cal F}\left(T_S\right) &=&  {\cal F}\left(T_0\right)
 + \frac{e^2}{2}\frac{r_Lr_R}{r_L+r_R}\big[ V_{1}I_{\rm det}(V_{1},T_{\rm det})
\nonumber\\ &&
+\,V_{2}I_{\rm inj}(V_{2},T_{\rm inj})
+P_{\rm det}(V_{1},T_{\rm det},T_S)
\nonumber\\ &&
+\,P_{\rm inj}(V_{2},T_{\rm inj},T_S) \big],
\label{TS}
\end{eqnarray}
where the function ${\cal F}(T)$ is defined as
\begin{eqnarray}
&& {\cal F}(T)=\int_0^T dT'\bigg\{  2T'F\left(\frac{\Delta(T')}{T'}\right)
\nonumber\\ &&
+\,2\Delta(T')\ln\left(1+e^{-\Delta(T')/T'}\right)
+\frac{\Delta^2(T')}{T'\left(1+e^{\Delta(T')/T'}\right)}\bigg\},
\nonumber\\
\end{eqnarray}
and $\Delta(T')$ denotes the standard BCS temperature dependence of the superconducting gap.

Equations (\ref{T1}), (\ref{T2}) and (\ref{TS}) constitute a complete system which allows one to determine 
the temperatures $T_{\rm inj}$, $T_{\rm det}$ and $T_S$ as functions of the bias voltages. This system of 
equations was resolved numerically by iterations. The corresponding results are compared to the experiments in 
Figs. \ref{fig_parameters} and \ref{fig_nonlocal} and discussed below in the next section.

\section{Discussion and conclusions}

The dependence of temperature on the bias voltage $V_{\rm inj}$ at $V_{\rm det}=0$ for the parameters of the sample A
predicted by the heating model is compared to the experimental data in Figs. \ref{fig_nonlocal}(b). In agreement with the experiment one observes that the superconducting wire is overheated stronger than the normal wire, in particular at low bias values. This effect can easily be understood since in this regime the heat conductivity of the superconductor (\ref{kappa})
is exponentially suppressed by the factor $\sim \exp[-\Delta/T_S]$.
Note that in our numerical simulations the broadening parameter $\Gamma$ was set equal to zero in the whole range of bias voltages.
Enhanced smearing of the $I-V$ curves encountered at high bias voltages
results from additional heating of the wires by the injector junction.

Our main results are depicted in Figs. \ref{fig_nonlocal}(c) and (d), where the predicted nonlocal conductances of the samples A and B are plotted. 
We observe that our model  not only qualitatively captures the behavior of $G_{\rm nl}$ as a function of the bias voltages $V_{\rm det}$ and 
$V_{\rm inj}$ but also correctly predicts the magnitude of the non-local conductance. It is also important to stress that a good agreement between 
theory and experiment was achieved with no fit parameters as all resistances and other parameters were measured independently. Hence, we conclude 
that strong non-local response observed in our samples at not very small bias voltages is indeed due to the effect of heating.

Note that the model prediction for the dependence of the non-local signal on the distance between the junctions turns out to be not very accurate, see Fig. 5b. 
We speculate that the main cause for this discrepancy might be the effect of a finite electron-electron relaxation length which was considered short
in our calculation. In any case, in order to quantitatively reproduce the two scale decay of the signal observed in our experiment 
it appears necessary to further refine the model employed in our theoretical analysis. 

In Figs. \ref{fig_nonlocal}(e) and (f), we also show the Coulomb correction predicted by eq. (\ref{dIdVCol}). Here, we have used the maximum of the charge imbalance signal measured at $V_\mathrm{det}=0$ as $G_{\rm nl}^{(0)}$, and the environmental resistance $R^E$ obtained from fitting the local Coulomb dip in the normal state. As can be seen, the Coulomb correction is qualitatively similar to the measured data, but too small by about three orders of magnitude.

Thermoelectric effects caused by the combination of a thermal gradient and a supercurrent have been observed in the 1970s.\cite{clarke1979b,pethick1979b} These might also contribute to the nonlocal effects reported here. For aluminum, the magnitude of the nonlocal voltage due to these effects was found experimentally \cite{heidel1981} to be
\begin{equation}
 \frac{V_\mathrm{det} G_\mathrm{det}}{j_\mathrm{s}\nabla T} \approx 10^{-21}~\mathrm{\frac{\Omega m^3}{K}}.\label{eqn_TE_estimate}
\end{equation}
For our experiment, we can estimate this contribution by assuming that the entire current, which is initially injected as quasiparticle current, is eventually converted to supercurrent. This yields an upper limit of the supercurrent density $j_\mathrm{s}\lesssim I_\mathrm{inj}/A$, where $A\approx 10^{-14}~\mathrm{m^2}$ is the cross-section area of the aluminum wire. Since the quasiparticle temperature is increased by about $1~\mathrm{K}$ over the bath temperature, and the observed effects decay on the length scale of a few microns, we can further estimate $\nabla T\lesssim 1~\mathrm{K}/\mathrm{\mu m}$. For comparison with our nonlocal conductance experiment, we express eq. (\ref{eqn_TE_estimate}) in terms of injector voltage and detector current and obtain
\begin{equation}
 \frac{I_\mathrm{det}}{V_\mathrm{inj}} \approx \frac{G_\mathrm{T}^2 \nabla T}{A}\times 10^{-21}~\mathrm{\frac{\Omega m^3}{K}} \lesssim 1~\mathrm{\mu S}.\label{eqn_TE_estimate_G}
\end{equation}
This is orders of magnitude smaller than the observed effects. Also, in our experiment the driving force of the thermal gradient is heating due to the injector bias. The heating power, and therefore $\nabla T$, is an even function of bias. The nonlocal conductance for this mechanism should also be even in bias according to eq. (\ref{eqn_TE_estimate_G}). We conclude that both by symmetry and order of magnitude the observed effects are not caused by charge imbalance in the presence of supercurrents and thermal gradients.

In summary, we demonstrated that heating can play a major role
dominating the non-local properties of three-terminal hybrid proximity structures at not very small bias voltages. In simple terms this effect can be understood as follows. Increasing the bias voltage in the injector
one effectively heats the superconductor which, in turn, yields
the temperature increase in the detector wire. As a result, the detector current changes thus providing the non-local response.
It turns out that in our samples this simple mechanism prevails -- at least at substantial bias voltages -- over more standard charge transfer mechanisms, such as charge imbalance or crossed Andreev reflection.
Quite generally, the heating strength is controlled by the ratio between the wire resistances and those of the tunnel junctions.
Heating effects are negligible provided this ratio is small, i.e. the junctions are more resistive than the wires. On the other hand, in the opposite limit of highly resistive wires heating gains importance and
essentially influences the system behavior.

In particular, in a typical beam-splitter setup with equal bias across both junctions, increasing the current in one branch will lead to an increase in the other branch as well. This mimics Cooper pair splitting in multi-terminal proximity devices, and an adequate analysis of the experimental data is needed in order to avoid possible misinterpretations. Our theoretical model provides a proper tool for such analysis.

Finally, we would like to point out that even at rather small voltages
heating effects in our structures can be non-negligible and should be treated on equal footing with, e.g., the effects of electron-electron interactions. This subject, however, requires a separate consideration which goes beyond the simple analysis presented here.

\section*{Acknowledgements}

This work was partially supported by the Deutsche Forschungsgemeinschaft within the Center for Functional Nanostructures and by RFBR grant No. 12-02-00520-a.

\appendix

\section{Coulomb correction to the current}

The Coulomb correction to the current can be derived from the theory of environmental Coulomb blockade \cite{ingold1992}.
For simplicity, we first consider a single detector tunnel junction between normal and superconducting bulk leads.
In this case the theory\cite{ingold1992} predicts the current in the form
\begin{eqnarray}
&& I_{\rm det} = -\frac{G_T^{\rm det}}{e}\int dEdE' \nu(E')
\nonumber\\ && \times\,
\big\{ f_{\rm det}(E,T_{\rm det})\big[ 1-f_S(E,T^S_{\rm det}) \big]P_{\rm det}(E-E'+eV_1)
\nonumber\\ &&
-\,\big[1-f_{\rm det}(E,T_{\rm det})\big] f_S(E,T^S_{\rm det}) P_{\rm det}(E'-E-eV_1)\big\},
\nonumber\\
\label{Idet}
\end{eqnarray}
where
\begin{eqnarray}
P_{\rm det}(E) = \int\frac{dt}{2\pi}\,e^{J_{\rm det}(t)+iEt}
\end{eqnarray}
is the probability to emit a photon with energy $E$ to the electromagnetic environment of the junction
defined in terms of the phase correlation function $J(t)$
\begin{eqnarray}
J_{\rm det}(t)&=&\frac{e^2}{\pi}\int_0^\infty d\omega\frac{{\rm Re}\,[Z_{\rm det}^E(\omega)]}{\omega}
\big\{(\cos\omega t - 1)\coth\frac{\omega}{2T}
\nonumber\\ &&
-\,i\sin\omega t\big\}.
\end{eqnarray}
Here $Z^E_{\rm det}(\omega)$ is the impedance of the environment "seen" by the detector tunnel junction. Here we will choose it in the form
\begin{eqnarray}
Z_{\rm det}^E(\omega)=\frac{R^E_{\rm det}}{1-i\omega\tau_{RC}},
\end{eqnarray}
where $R^E_{\rm det}$ is an effective Ohmic shunt of the detector junction and $\tau_{RC}\approx R^E_{\rm det}C_{\rm det}$
is a (short) charge relaxation time depending on the effective junction capacitance $C_{\rm det}$. With a reasonable
accuracy one can identify the shunt resistance $R^E_{\rm det}$ with the resistance of the
normal wire attached to the detector junction $r_{\rm det}^N$, i.e. $R^E_{\rm det}\approx r_{\rm det}^N$.

Here we are mostly interested in the non-local contribution to the detector current.
It can be derived in the same way as the main contribution to the current (\ref{Idet})
repeating the procedure outlined, e.g., in the review \cite{ingold1992}.

As a first step one assigns
a phase factor $e^{i\hat\varphi_j(t)}$ ($j=$inj or $j=$det)
to the tunneling amplitudes of the junctions $t_j$ treating the phases $\hat\varphi_j(t)$
as quantum operators. This phase is related to the voltage fluctuations across the junctions,
$\dot{\hat\varphi}_j(t) = e\delta\hat V(t)$. Next, one performs the standard perturbative expansion
of the current in powers of the tunneling  Hamiltonians of the two junctions to the lowest
non-vanishing order $\propto t^2_{\rm inj}t^2_{\rm det} e^{i\hat\varphi_{\rm inj}(t_1)}e^{-i\hat\varphi_{\rm inj}(t_2)}
e^{i\hat\varphi_{\rm det}(t_3)}e^{-i\hat\varphi_{\rm det}(t_4)}$. Subsequent averaging over the phase fluctuations
results in the product of the two functions $P_{\rm inj}(E_1)P_{\rm det}(E_2)$. Leaving out further technical details, we go over to the final result which reads
\begin{eqnarray}
&& \delta I_{\rm det}^{\rm nl} = -\frac{G_{\rm nl}^{(0)}}{2e}\int dE dE_1dE_2 \theta(|E|-\Delta)
P_{\rm inj}(E_1)P_{\rm det}(E_2)
\nonumber\\ && \times\,
\big[ f_{\rm det}(E-E_2+eV_1) + 1 - f_{\rm det}(E+E_2+eV_1) \big]
\nonumber\\ &&\times\,
\bigg\{
\left( 2 - \frac{\Delta^2}{E^2-\Delta^2} \right)\big[ f_S(E,T^S_{\rm inj})\big( 1-f_{\rm inj}(E+E_1+eV_2) \big)
\nonumber\\ &&
-\, \big( 1-f_S(E,T^S_{\rm inj})\big)f_{\rm inj}(E-E_1+eV_2) \big]
\nonumber\\ &&
-\, \frac{\Delta^2}{E^2-\Delta^2} \big[ f_S(E,T^S_{\rm inj})\big( 1-f_{\rm inj}(E-E_1-eV_2) \big)
\nonumber\\ &&
-\, \big( 1-f_S(E,T^S_{\rm inj})\big)f_{\rm inj}(E+E_1-eV_2) \big]
\bigg\}.
\label{dIdet}
\end{eqnarray}
Bearing in mind the property of the Fermi function $f(-E)=1-f(E)$, it is straightforward
to check that in the non-interacting limit, where $P_{\rm inj}(E)=P_{\rm det}(E)=\delta(E)$,
the correction (\ref{dIdet}) reduces to the charge imbalance correction defined in the
Eq. (\ref{Inl}).

In the relevant for our experiment weak Coulomb blockade limit $g^E_{j}\equiv 2\pi/e^2R^E_{j} \gg 1$ we may express the functions $P_j(E)$ in the form
\begin{eqnarray}
P_j(E) = \delta(E) + \delta P_j(E),
\end{eqnarray}
where we defined
\begin{eqnarray}
\delta P_j(E) &\approx & \frac{2}{g^E_{j}}\frac{\left(e^{-\gamma}E\tau_{RC}\right)^{2/g^E_{j}}}{E\left(1+E^2\tau_{RC}^2\right)\left(1-e^{-E/T}\right)}
\nonumber\\ &&
-\frac{2}{g^E_{j}} \delta(E)\int  \frac{dE'\,\left(e^{-\gamma}E'\tau_{RC}\right)^{2/g^E_{j}}}{E'\left(1+(E')^2\tau_{RC}^2\right)\left(1-e^{-E'/T}\right)}.
\nonumber
\end{eqnarray}
Within this approximation and in the limit $T_{\rm inj},T_{\rm det},T_{\rm inj}^S,T_{\rm det}^S\to 0$
the correction to the non-local conductance derived from the general expression (\ref{dIdet}) reduces to the form (\ref{dIdVCol}).

\bibliography{lit.bib}

\end{document}